\documentstyle[12pt]{article}
\begin{document}

\title{Measure of irregularity for Dirac, Schwinger,
       Wu-Yang's bases in the~Abelian monopole theory
       and affecting of the gauge symmetry principle by
       allowance of singularity in physics}

\author {V.M.Red'kov \\
     Institute of Physics, Belarus Academy of Sciences\\
F Skoryna  Avenue 68, Minsk 72, Republic ob Belarus\\
e-mail: redkov@dragon.bas-net.by}

\maketitle

\begin{abstract}.
In the literature concerning the~monopole matter,
three gauges: Dirac, Schwinger, and Wu-Yang's,
have been contrasted to each other, and the~Wu-Yang's often appears as
the~most preferable one.   The~article aims to analyse this view by
interpreting the~monopole situation in terms of the~conventioal Fourier
series theory; in particular, having relied on the~eminent Dirichlet
theorem.  It is shown that the~monopole case can be  labelled as
a~very spesific and even rather simple class of problems in the~frame
of that theory: all the three monopole gauges amount to practicaly
the~same one-dimentional problem
for functions given on the~interval $[0, \pi]$, having a~single point
of discontinuity; these three vary only in its location.

Some general aspects of the Aharonov-Bohm effect are discussed; also
the way of how any singular potentials such as
monopole's, being allowed in physics,   touche the
essence of the physical gauge principle itself is considered.
\end{abstract}

\newpage
\subsection*{1. Introduction}

The study of monopoles has now reached a point where  further
progress depends on a~clearer  understanding of this  object  that
had been  available  so  far.  As  evidenced  even  by  a  cursory
examination of some popular surveys [1,2],   the whole monopole area
covers and touches quite a~variety  of  fundamental  problems.  In
particular, following the~original and brilliant pattern  given by
Dirac [3,4],  physicists  always   were  especially  concerned  with
relevant singularity  problems.  Besides,   throughout  all
the~history of this matter,   conceiving  itself  of  an~idea  of
monopoles has been always associated with concept of singularity.
Leaving aside a~major part  of various  monopole  problems,   much
more comprehensive area in itself, just  those  singularity aspects,
notoriously known and generally accepted as difficult, and what is
more, hitherto conclusively unsettled,  will be  a~subject  of
the~present study\footnote{Though evidently,  ultimate  answers  have  not
been found by this work as well,  it  might be hoped that a~certain
exploration into and clearing up  this matter have been achieved.}.

In the  work, only  the~monopole's singularities are discussed.
In this connection, it should be emphasized at once  that though much
more involved irregular (even not monopole-like  ones)  configurations
are consistently invented and reported in
the~literature; in the~same  time, it might be hoped, that
just above-mentionned, old and familiar,  monopole-based peculiarities
came  to  light, again  and  repeatedly, in a~somewhat disguised form, when
considering those generalized systems. So, in the~light of that connection,
a~more particular situation, investigated in the~paper, is of reasonable
interest for the~more large number of problems.  In any case,
as evidenced by all the~history of study of monopoles, even this
seeminly plain, at first glance, case has turned  out to  be
far too formidable an~undertaking theoretically (and all the  more
experimentally).

Once the monopole had been brought into scientific  usage  at
the~quantum  level,  its main singular properties had been noted
and examined. The~background of thinking
the~whole  monopole problem in that time can easily be traced; it was
obviously tied up with the~most outstanding  point  of  hypothesis
about a~magnetic  charge ($g$):  the  Dirac's  electric  charge
quantization  condition.  Just   the~latter  was the~first
consideration in any  assessment  of  the  problem  in~a~whole.
Moreover, this  quantization  condition  had  occurred  from
the~Dirac's attempts to get over some  difficulties  concerning the basic
requirements of continuity  in  quantum  mechanics,  i.e.  in
the~process of solving again the~same singularity problems.

Also,  the  Schwinger's  attempts  to  dispose  of   magnetic
charge's singularities and modify the quantization condition were
of significant implication to subsequent  discussing the~monopole
matter. In particular, his seminar paper [5] brought  out a~sharp
separation of characteristics of integral and half-integral $eg$ cases
($e$ and $g$ denote, respectively,  electric and magnetic charge) and based
the full discussion on a~study of such peculiarities.

In  essence,  the~more  recent,  and  of  great   popularity
currently, approach by Wu and Yang [6] adheres closely  to  the~same
Dirac and Schwinger's regard for the~importance of continuity
requirements in presence of the monopole and  for  the~importance  of
establishing  some reasonable and intelligible rules for  handling
all singularities encountered. They (Wu and Yang) renewed the~old
Dirac's arguments, essentialy  updated the~relevant  mathematical
techniques, and finally invented, in~a~sense, a~new  mathematical
and physical object; the~latter is designated now often as  the~Wu-Yang
monopole\footnote{Their   approach seemingly  enables us to  surmount
the~old problem of monopole singularities; though, as may be noticed
in more close investigating (see below), it does not explain away all
its concomitant doubts and obscurities.}.  The~crucial moment in
their contemplating the~problem of monopole peculiarities had been the~same
old intention to overcome all  the~singularities  occurred.  Starting
from the~observation that the~Schwnger's  potential is  not  well
difined at the $x_{3}$-axis; as well as the~Dirac's potential is
undetermined at a half-axis (it is either
$x^{+}_{3}$   or   $x^{-}_{3}$ ), Wu and  Yang  had
suggested  a~simple  trick:   instead   of   a~globally   given
electromagnetic  4-potential
 $ \{ \; A_{\beta }(t,\vec{x}) , \;  \vec{x} \in  R_{3} \; \}$
(in particular, the~Dirac or  Schwinger's  monopole potentials  had
been meant), they had said, one could  use a~pair of non globally
given ones, which consists of  two  Dirac's  type  sub-potentials
$\vec{A}^{(N)}$  and   $\vec{A}^{(S)}$
given respectively  in   two  half-spaces (just  in
their own regions with no singularities)  completed each other up to
the~whole 3-space as follows
               $\vec{A}_{WY} \; = \; \{ \;
             \vec{A}^{(N)}  \; \bigcup \;  \vec{A}^{(S)}\; \} $;
so that $\vec{A}^{(N)}$ has no singularity at the~positive
 half-axis $x^{+}_{3}$ as well as $\vec{A}^{(S)}$ does not have any
singularity at the negative one $x^{-}_{3}$.

Thus, as  often asserted, the~absence  of  singularity,  at
least locally, had been achieved and thereby the clouds over this
part of the subject had been dispersed. Therefore,  the crisis  in
the scientific picture of this  matter  had  been  set  out  in  a
seemingly perfect fashion, thereby  obviating any further doubts.
Such a~local  charts-based  approach  to  this  and  a~variety
of similar situations has been extensively and in great  detail
elaborated, so as an~absolutely new  mathematical language  and
physical methodology have  been worked out to date.  And  now,  it
is almost generally accepted outlook to this matter  that  such
a~locally  achieved  continuity  provides  us  with  a~substantial
progress  in   studying  and  understanding   any systems
containing some singularities.   This is where the~subject  stands
now --- very roughly speaking, of course\footnote{It would carry us very
far afield to discuss at any lenghth such  purely mathematical
considerations; instead, the working language
of the paper is going to be much more conventional, intuitive, and
physically felt.}.

The~aim of the~present work, in particular, is to demonstrate that there
exist some grounds to query  whether  the~monopole singularities  have  been
ruled out indeed; the~article suggests that such an~outlok  hardly
would stand close examination\footnote{So, the~tranquillity  dominating
among majority of physisists on this~problem  is not  justified
anyhow. To avoid any misunderstanding, it must be emphasized at
once that this  work is in no way a~strenuous  objection against
the~Wu-Yang formalism and its concomitant methodology. Also, author does
not claim that the method by Wu and Yang is mistaken or misleading anyhow;
instead, the~article  just  points certain inherent features  which
delimit its  powers to  some  natural bounds, and puts forward
a~possible developement  complementary to it.}.

So, our further work is laid out as follows. Sec.2 treats, in
a~fairly unusual way,  an~indeterminated character of the~above
potentials. It is convenient first to discuss in detail one  gauge
--- for definiteness we start with  the  Schwinger's; the~considering
of two others is deferred to Sec.3. In so doing, a~special  notice
is given to comparison of the~representations  of  the~ Schwinger's
monopole  in Cartesian and spherical coordinates; at this we trace
a~delicate cancellation between  different terms in
the~process of this coordinate change. The~spherical picture  is
treated as preferable to Cartesian one;  the~reasons to this are  that
in spherical basis  a~major part (though not the~most essential one)
of  singular  manifestations  of the~monopole's $4$-potential
is hidden (effectively)
by known $\theta ,\phi $-coordinates' singularity. As shown, through
the~use of the conventional generaly-covariant  tensor formalism,
the~monopole singularity problem is reduced to a~single function
$f(\theta ) \; ( A_{\phi } = g \cos  \theta  )$
given on the~interval $\;[0 , \pi  ]\;$. In this connection, one ought
to keep in mind the~known (and apparently hidden) indeterminacy at
the~axis $x_{3}$ for the~spherical  vector   $\vec{e}_{\phi }$.
For the~case under consideration, this circumstance implies that
to any non-singular
physical situation there must correspond a~function
$A_{\phi }^{regul.} = f(\theta ) ^{regul.} \;$ with zero-boundary conditions
at these points $\theta = 0 \; , \; \pi $.

In Sec.3, the other two gauges  (Dirac  and Wu-Yang's)  are
looked at, of course on the~line used in Sec.~2. Then,  the~prime
question is that concerns the~hierarchy (if  any)  among  three of them;
i. e. --- whether or not these three gauges are  unequally singular ones.

To produce any constructive and justified criterion for counting
a~{\em quantity of singularity}, we contrast the~above three functions
($A^{S}_{\phi}, A^{D}_{\phi}, A^{WY}_{\phi}$) and
accompaning boundery conditions (for definiiteness, here the Schwinger's
case is taken)
$$
 f^{S}(\theta ) = A^{S}_{\phi } = g \cos  \theta  \; : \qquad
\theta \in [ 0 , \pi ] \; , \;\; f^{S}(0) = +1 \; , \;\; f^{S}(\pi ) = -1
\eqno(1.1)
$$

\noindent with  their  counterpart in absence of any singularity  at
the~axis $x_{3}$; namely
$$
 f^{0} (\theta ) = A^{0}_{\phi } \; : \qquad
\theta \in [\; 0 \;,\; \pi\; ]\; , \;\; f^{0}(0) = 0 \; ,
\;\; f^{0}(\pi ) = 0 \;  .
\eqno(1.2)
$$

\noindent This may be expressed as follows: while a~non-singular problem
being  associated with a~definite case in the~frame of the~Fourier series
analysis, for  which  the~boundery conditions are specified as null
ones,  the~monopole problem should be referred to its own type of Fourier
problem.
Definitely, all the differences concern and come from variations in
boundary conditions and continuity properties, which either remain the same or
get violated\footnote{So, merely the~reformulation of the~monopole problem in
other terms unables us to make good use for the~conventional mathematical
theory of Fourier series in studying the~monopole problem. In particular,
as a~general basic point representing this theory and obviously touching
the~problem under consideration,  the~well-known Dirichlet theorem has been
taken.}.
To formalize mathematically this observation (see (1.1) and (1.2)),
we have determinated   a~quantity  (designated  by $\mu_{inv.}(\vec{A})$
which  might be  treated  as  a~measure  of   singularity   for
electromagnetic potential  $\vec{A}$. Besides, and what is more, we show
that  this  $\mu_{inv.} $ has the same one value for all the three monopole
potentials:
$$
\mu_{inv.} (\vec{A} ^{S}) = \mu_{inv.} (\vec{A} ^{D}) =
 \mu_{inv.} (\vec{A}^{WY}) .
$$

\noindent Therefore, in that sense, all three gauges amount to each
other and there are no reasons to prefer any  one  of  them\footnote{This
contrasts somewhat  with  a~common viewpoint in the~literature, when
the~Wu-Yang approach to  monopole problem has been regarded often as having
some advantage.}.
Extending this observation, it is reasonable  to  conjecture  that
the~$\mu (\vec{A})$  is gauge-invariant quantity, i.e. it  will not change
when  we perform an~arbitrary $U(1)$ gauge  transformation  with
any type of singulariry involved.

In addition, else one type of measure of singularity of electromagnetic
potential (it called an `additive' measure  $\mu_{addit.} (\vec{A})$)
has been introduced. In contrast to $\mu_{inv.}$ the latter  must
substantially vary  when any piecewise continuous (in the sense of functions
of spatial coordinates)  gauge transformations are used; for more
detail see below in Sec.4).

In sec.5, in terms of those measures   $\mu_{inv.}$ and
$\mu_{addit.} (\vec{A})$, we consider  several aspects
of the~Aharonov-Bohm effect and discuss some inherent requirements
implied by the~conventional gauge principle.
Particularly, we take special notice of
the fact that any singular potentials such as
monopole's, being allowed in physics,  significantly  touche the
essence of the physical gauge principle itself.

\subsection*{2.  Schwinger's potential in Cartesian and
                 spherical coordinates}

First let us consider more closely some facts on  Schwinger's
gauge which are to be counted on  in the~following. As well known,
the~Schwinger's potential [5] is given by
$$
\vec{A}^{S}(x) = - \; g \; { [ \; \vec{r} \times  \vec{n}\; ] \;
  ( \vec{r} \; \vec{n})  \over  r \; ( \; r^{2} \; - \;
(\vec{r} \; \vec{n})^{2}\;   )}
\eqno(2.1)
$$

\noindent where $\vec{n}$   stands  for an~arbitrary 3-vector and
thereby it represents an~additional parametre fixing a~certain geomerical
orientation of the~monopole in the 3-space.  At once, it should be noted
that this potential $\vec{A}^{S}(x)$ is not  a~well  defined  quantity at
the~whole $x_{3}$-axis; it is only  a~$(0/0)$-expresion when
$\vec{r} = \pm\; r \; \vec{n}$.

Setting $\vec{n} = (0, 0, +1)$ and translating this $\vec{A}_{S}$ to
the~usual spheric  coordinates, one  gets
$$
A^{S}_{\phi} = \; g \; \cos \theta  \; .
\eqno(2.2)
$$

\noindent The $A^{S}_{\phi }$  has non-vanishing values at $\theta  = 0$ and
 $\theta  = \pi $:
$$
 A^{S}_{\phi } = +\; g \;\;  if \;\;   \theta  = 0    \qquad   and \;\;
                 -\; g \;\;  if \;\;   \theta  = \pi  \; .
$$

\noindent It is the point to remember that  as $\theta = 0$ or $\theta = \pi$,
then the~basis spherical vector $\vec{e}_{\phi }$ has no single  sense:
there exists a~set  of  possibilities for $\vec{e}_{\phi }$   rather than
only one that. This circumstance obviously  comes from an~original
indeterminacy of the~spherical coordinate $\phi $  at the $x_{3}$-axis.
For this reason, a~genuine sense of $A_{\phi }$  at  the~axis $x_{3}$
should constitute just a~characterization of $A_{\phi }$ in a neighborhood
of this axis rather than any specific values for it at the~points
lying in the~axis $x_{3}$. In other words, the~potential from (2.2) provides us
with a~non-single-valued function of spatial points just at the~axis
$x_{3}$.

Evidently,  the~above peculiarities of the~monopole potential (2.2)
do not  originate in  the~irregularity properties of the~spherical
coordinates $(\theta,\phi)$.
Indeed, some discontinuity occurs likewise in the~Cartesian coordinates,
when  the~monopole potential is described by (2.1); there it exhibits
its own indeterminacy of the~$(0/0)$-kind  at the~axis $x_{3}$.
Let us  look at this more closely.

Because of the~potential  $\vec{A}_{S}$ from (2.1) is formally meaningless at
the~axis $x_{3}$,  we
should  look  at its values in  the~adjoining neighborhood defined by
$$
\vec{r} =   \; (0,\; 0, \;z ) \; + \;
\epsilon \; (m_{1}, m_{2}, m_{3})\; = \;
 \vec{z} + \epsilon \;  \vec{m} \; , \qquad
\left ( \; \vec{m}^{2} = 1 \; , \;  m_{3} \neq  \pm  1 \; ,
\;  \epsilon  \rightarrow  0  \; \right ) .
$$

\noindent So,  as a~representative of the~monopole potential at  $x_{3}$,  one
has the~quantity depending additionally on the~vector $\vec{m}$:
$$
\vec{A}^{S}(\vec{z}, \; \vec{m}) \; \equiv \;
\lim_{ \epsilon \rightarrow 0} \; \vec{A}^{S}(\vec{z} \; +
\;  \epsilon \; \vec{m} )
$$

\noindent that is
$$
\vec{A}^{S}(\vec{z}, \; \vec{m}) \; = \;
\lim_{ \epsilon \rightarrow  0} \;  \left [ \;  - g \;
{ \epsilon \; (\; m_{2} \; \vec{e}_{1} \; - \; m_{1} \; \vec{e}_{2}\;) \;
(\;z \; + \;  \epsilon \; m_{3} \;)
\over
\sqrt { \epsilon ^{2} \; m^{2}_{1} + \epsilon ^{2} \; m_{2} \; +
(z + \epsilon \;  m_{3})^{2} } \;\; \epsilon ^{2} \;
( m^{2}_{1} + m^{2}_{2} ) } \; \right ]
$$

\noindent where $\vec{e}_{i}$ denotes the usual Cartesian orthonormal vectors.
Further we have to draw  distinction  between $z = 0$   and $z \neq  0$.
First,
$$
z \neq 0 \; :   \qquad \vec{A}^{S.} (\vec{z}, \vec{m} ) = \;
-\; \infty \;  g \; Sgn \; z
 \left ( \; { m_{2} \; \vec{e}_{1} - m_{1} \;\vec{e}_{2} \over
m^{2}_{1} \; + \; m^{2}_{2} }\; \right )
\;\; where \;\; \infty = \lim_{} \; { 1 \over  \epsilon} \; \;
 as \; \epsilon \rightarrow  0 \; .
\eqno(2.3a)
$$

\noindent The unit vector $\vec{m}$ can be characterized  by
$$
 m_{1} = \sin  \Theta  \cos \phi  , \;\;
 m_{2} = \sin  \Theta  \sin \phi  , \;\; m_{3} = \cos \Theta
$$

\noindent where the quantity  $\Theta $ does not coincide with
the~spatial coordinate variable $\theta $, whereas the $\phi$ is
the~usual spherical coordinate.  In a~sequence,  the~above vector
$\vec{A}^{S}(\vec{z}, \vec{m})$  can be reexpressed  as
$$
\vec{A}^{S}(\vec{z} , \vec{m} ) = \; g \; sgn \; z \;
 { \infty \over \sin \Theta } \;\;
\vec{e}_{\phi }
\eqno(2.3b)
$$

\noindent where $\vec{e}_{\phi }$  designates a~combination
$( \vec{e}_{\phi } = \sin \phi \; \vec{e}_{1}  -
\cos \phi \; \vec{e}_{2} )$.
It should be noted  that  the  factor $\sin ^{-1}\Theta$  (in  the
(2.3b))  is not essential one in the~sense that
the~symbol $\infty$  (having remembered  that  $\Theta \neq  0, \pi  )$  is
presented there as well. So, instead of (2.3b) one may write down
$$
\vec{A}^{S}(\vec{z} , \vec{m} ) = \; g \; sgn \; z \;
  \infty \; \vec{e}_{\phi } \; .
\eqno(2.3c)
$$

In turn, for the $z = 0$  case one produces
$$
\vec{A}^{S}(\vec{r}=0, \; \vec{m} ) = \;\;  - g \;\;
{ \infty \over \sqrt{m^{2}_{1} + m^{2}_{2} }}
\;\;  m_{3} \;\; { m_{2} \; \vec{e}_{1} - m_{1} \; \vec{e}_{2} \over
\sqrt{m^{2}_{1} + m^{2}_{2} }}
\eqno(2.4a)
$$

\noindent or further
$$
\vec{A}^{S}(\vec{r}= 0, \;\vec{m} ) = +\; g\; \; { \infty  \over \sin \theta }
\; \;\cos \theta \; \vec{e}_{\phi } \; \sim + \; g\; \infty
\;\cos \theta \; \vec{e}_{\phi }
\eqno(2.4b)
$$

Evidently, contrasting that {\em Cartesian} representation (see (2.3) and (2.4))
with an~alternative {\em spherical}  one
$$
A^{S}_{\phi }( \vec{r} = \vec{z}) = g \; sgn \; z  , \qquad
A^{S}_{\phi }(\vec{r} = 0  ) = g \cos  \theta \;
\eqno(2.5)
$$

\noindent we conclude that the {\em spherical} description seems
formally  a bit less singular than Cartesian's: the $\infty$ is absent in
{\em spherical} picture.  In other words, the~singularity properties
of the monopole at the~axis $x_{3}$ fall naturally into two
groups, one of which (the $\infty$ is meant) is subject to an~incidental
coordinate choice and another one (the factor $g \; \cos  \theta$)   reflects
a~properly monopole's essence. So, it seems that
just the~factor $g \cos \theta $  carries a~monopole  quality  after
leaving out the~all  complications  originating  in
the Cartesian coordinate system\footnote{It is somewhat surprising that so
simple function as $g \cos\theta $  tells us a~lot  about  the~monopole  and
contains potentially a~great deal of  information  concerning
the~magnetic charge.}.

It is of primary significance to the further exposition, that
the Schwinger gauge exhibits a~singularity  both in
the~positive and negative half-axes $x_{3}$, as well as in  the~zero-point
(0,0,0). One should repeat again: these  singularities  consist
solely in the fact that the~values of the~monopole  potential  at
the~axis $x_{3}$ are a~function of spatial directions  that  characterize
possible ways of approaching these points $(0, 0, z)$.
The same  may be expressed as assertion that  the~monopole
potential provides us with an example of quantity which is  not
a~single-valued function of spatial points at  the~axis $x_{3}$.

All points of the positive half-axis $x_{3}$   are  exactly  alike
with respect to  their discontinuity properties. Therefore, as
a~possible method to  describe this,  one may try
the~following formulation: the  half-axis $x^{+}_{3}$    provides $2\pi $
directions of discontinuity. A~completely  analogous
statement concerns the~negative half-axes $x^{-}_{3}$. Finally,
the~null point gives us $(2\pi  \otimes  \pi )$  irregularity  directions.
Thus,  the~Schwinger monopole potential, in~a~whole, can be schematically
sketched by
$$
A^{S}_{\phi } \;\; \rightarrow \;\;
\left \{ \begin{array}{llc}
        x^{+}_{3} \;  &\sim \; &  ( + 2 \pi  )\; g  \\
        0         \;  &\sim \; &  ( + 2 \pi  \otimes  \pi  ) \; g  \\
        x^{-}_{3} \;  &\sim \; &  ( - 2 \pi  ) \;  g
\end{array} \right.
\eqno(2.6)
$$

\noindent where the signs  "$+$"  and  " $-$"  serve to remind  us of
the $sgn(z) \;$  in the $A_{\phi }(z) = g  \; sgn(z)\;$ ; just
the~function $sgn(z)\;$  leads us to distinguish the $x^{+}_{3}$
 and $x^{-}_{3}$   half-axes when  characterizing  the~monopole singularities.

\subsection*{3. The Dirac and  Wu-Yang's representations}

Now, from the same point of view, we are to analyze Cartesian
and spherical pictures for the~Dirac gauge. The Dirac potential is as follows
(in the following, let  $\vec{n}$ be equal $(0, 0, +1)$)
$$
\vec{A}^{D(+)}\; =  g\; {[\; \vec{n} \times \vec{r}\; ] \over r \;
(\; r\; + \; \vec{r} \; \vec{n}\; )}\; = \;
 \; g \; { - x_{1} \; \vec{e}_{2} \; + \; x_{2} \; \vec{e}_{1}  \over
  r \; (\; r \; + \; x_{3} \;)} \;.
\eqno(3.1)
$$

\noindent In contrast to the Schwinger's case,  here  an~indeterminacy
$0/0$  is located  only at the~negative   half-axis $x^{-}_{3}$ as well as at
the~zero-point, while  the~$\vec{A}^{D(+)}$ has no  discontinuity at the~positive
one: $\vec{A}^{D(+)}(x^{+}_{3}) \equiv  0$.

Applying the~limiting procedure  above  to  the~$\vec{A}^{D(+)}$,
one easily  produces
$$
\vec{A}^{D(+)}( x^{-}_{3} ,\; \vec{m} ) = \infty \; ( - 2 g )\;
 ( m_{2} \; \vec{e}_{1} \; - \; m_{1} \; \vec{e}_{2}) \sim
\infty \; ( - 2 g )\; \vec{e}_{\phi}\;  ,
$$
$$
\vec{A}^{D(+)}(\vec{r}= 0, \; \vec{m} ) = \infty  \; g \;
\left (\; m_{3} \; - \; 1 \; \right ) \; \vec{e}_{\phi} \; .
\eqno(3.2a)
$$

In the {\em spherical} picture, the~Dirac potential is
$$
A^{D(+)}_{\phi } = \; g \; (\; \cos \theta  \; - \; 1\; )
$$

\noindent correspondingly, its singularities are characterized by
$$
A^{D(+)}_{\phi }(\vec{r} = 0 ) \; = \; g \; (\cos \theta - 1) \; , \qquad
A^{D(+)}_{\phi }( x^{-}_{3}, \; \vec{m} ) \; = \; - 2\; g    \; .
\eqno(3.2b)
$$

\noindent
It is the absence of discontinuity at the~positive  half-axis
$x^{+}_{3}$ (when $\vec{n} = ( 0, \; 0,\; +1)$)
that singles out the~Dirac gauge $\vec{A}^{D(+)}$.
In  comparison with the~Schwinger's that is singular both in the $x^{+}_{3}$
and $x^{-}_{3}$ half-axis, the~Dirac gauge seems less singular.

So, at first
glance, the  $D$-gauge  looks preferable to the  $S$-gauge; but on
closer examination we will see that it is hardly  so. In particularly,
this ({\em good} at the  $x^{+}_{3}$-axis)  gauge can be sketched by
(compare it with (2.6))
$$
A^{D(+)}_{\phi } \rightarrow
\left \{ \begin{array}{llc}
           x^{+}_{3}  \; & \sim \; & 0  \\
             0 \;        & \sim \; & ( + 2 \pi  \otimes  \pi ) \; g  \\
           x^{-}_{3}  \; & \sim \; & ( - 2 \pi  ) \; 2 g  \; .
\end{array} \right.
\eqno(3.2c)
$$

\noindent In the same time it should be noted that, by some
intuitive considerations, the~Dirac  gauge  appears
to be equivalent to  Schwinger's  because,   in~a~sense, the~Dirac
discontinuity at $x^{-}_{3}$  looks more intense  than  Schwinger's:
to realize this it suffices to take notice of  the~factor $2g$   at
(3.2c)  in contrast to the~factors: $+g$  and $-g$ in (2.2). This
would mean that  through the~transformation  $S$-gauge into
$D$-gauge  one has managed  to reduce the discontinuity set  from $\{
 \; x^{+}_{3} \oplus (0,0,0) \oplus x^{-}_{3} \; \}$ into $\{ \;
 (\;0,\; 0, \; 0\;) \oplus x^{-}_{3}\; \}$, but in the~same time one
has augmented a~{\em power} (or {\em intensity}) of  the~remaining
discontinuity set.

In addition to the~above, it should  be  reminded that  the~Dirac potential
(with $\vec{n}$ specified  as  $(0, 0, -1)$) is given by
$$
\vec{A}^{D(-)} = \; g \;
{[\; - \vec{n} \times  \vec{r}\; ] \over  r \;(\; r \;-
\;\vec{r} \; \vec{n}\; )} = - g
\; { - x_{1} \; \vec{e}_{2} \; + \; x_{2} \; \vec{e}_{1} \over
 r \; (\; r \; - \; x_{3}\;)}
\eqno(3.3a)
$$

\noindent and, in turn, it has a  $0/0$  indeterminacy  at  the~positive
half-axis $x^{+}_{3}$, which leads to
$$
\vec{A}^{D(-)}( x^{+}_{3},\; \vec{m}) \; = \;
\infty \; ( + 2 g ) \; (\; m_{2} \; \vec{e}_{1} \; - \; m_{1} \; \vec{e}_{2}\; )
\sim \; \infty \; ( + 2 g ) \; \vec{e}_{\phi} \; ,
$$
$$
\vec{A}^{D(-)} ( \vec{r} = 0, \;\vec{m}) =
 \infty \; g \;  ( m_{3} + 1 ) \; \vec{e}_{\phi} \; .
\eqno(3.3b)
$$

\noindent Instead of (3.2b)  now  we  have
$$
A^{D(-)}_{\phi } = g (\; \cos \theta  - 1\;) \;\rightarrow \;
\left \{ \begin{array}{lc}
          A^{D(-)}_{\phi } ( x^{+}_{3}, \vec{m} ) \; & =  \; + 2\; g \\
          A^{D(-)}_{\phi } ( 0 , \vec{m} )       \;  & = g (\cos \theta + 1) \; .
\end{array} \right.
\eqno(3.3c)
$$

Now, it is the point to introduce  the Wu-Yang  potential [6]. It
is determined by the following constituent form
$$
\vec{A}_{WY} =
\left \{ \begin{array}{l}
     \vec{A}^{(N)} = \vec{A}^{D(+)} \;\; if \;\;  0 \le  \theta  < \pi /2   \; ,\\
     \vec{A}^{(S)} = \vec{A}^{D(-)} \;\; if \;\;  \pi /2 < \theta  \le  \pi \; .
\end{array} \right.
\eqno(3.4)
$$

\noindent As evidenced by  its   definition,  this  potential
$\vec{A}_{WY}$  has  no discontinuity both in  the~ $x^{+}_{3}$
and $x^{-}_{3}$    half-axes. But, in author's opinion, it would be
untenable to  justify  preferable utilizing   the~latter  gauge only.
The~reason is that one should  give  special attention  to
the~following: in $(WY)$-gauge, some  discontinuity  occurs  at
 the~$(x_{1}- x_{2})$-plane and this must be taken into account.  One
should  remember that the~term `discontinuity' itself implies that there is,
at certain  points,   any dependence  on  possible  directions   of
approaching them; and just so the~Wu-Yang potential looks at
the~$(x_{1}- x_{2})$-plane:
$$
A^{WY}_{\phi} (\theta = \pi /2 + 0) \; = \;  g \; ( -1 ) \; , \qquad
A^{WY}_{\phi} (\theta = \pi /2 - 0) \; = \;  g \; ( +1 ) \; .
\eqno(3.5)
$$

\subsection*{4. Hierarchi among the~Dirac, Schwinger, Wu-Yang's
            gauges  and  measure of the~monopole irregularity}

The whole situation with~the monopole gauges (described in Sections 2 and 3)
can be reformulated and
summarized as  follows. Original  $S$-  and  $D$-gauges  provide  us  with
{\em strong} singularities concentrated along the~$x_{3}$-axis.
In going from the $S$- and $D$-gauges to the~Wu-Yang's  we  scatter
the points of discontinuity over the $(x_{1} - x_{2})$-plane, so that
this plane turns out to be filled up with irregularity  points. These
latter  are less singular that former ones  because these new irregularities
are  only two-valued ones; however, as a~way of compensation, the~number
of such irregular  points becomes much more greater.  So,  each of these
three gauges is equally singular, with  its  own character of discontinuity,
varying only in location. The case may be illustrated by the
following picture

\begin{center}
Fig.1
\end{center}
\vspace{5mm}
\unitlength=0.75 mm
\begin{picture}(170,140)(-20,0)
\special{em:linewidth 0.4pt}
\linethickness{0.4pt}

\put(+5,+140){$A^{z_{+}} _{\phi} = +g$}
\put(+5,+10){$A^{z_{-}} _{\phi} = -g$}

\put(+5,+75){\vector(+1,0){60}}   \put(+70,+70){$y$}
\put(+35,+10){\vector(0,+1){130}}   \put(+30,+135){$z$}
\put(+65,+105){\vector(-1,-1){60}}  \put(+5,+55){$x$}

\put(+35,+100){\circle*{1}}
\put(+50,+50){\vector(-1,0){12}}
\put(+20,+50){\vector(+1,0){12}}
\put(+45,+60){\vector(-1,-1){8}}
\put(+25,+40){\vector(+1,+1){8}}

\put(+35,+50){\circle*{2}}
\put(+50,+100){\vector(-1,0){12}}
\put(+20,+100){\vector(+1,0){12}}
\put(+45,+110){\vector(-1,-1){8}}
\put(+25,+90){\vector(+1,+1){8}}

\put(+90,+140){$z = +0$}
\put(+105,+120){$e_{\phi}$}
\put(+130,+140){$A^{WY}_{\phi}= -g$}
\put(+90,+40){\vector(+1,0){60}}   \put(+150,+35){$x$}
\put(+120,+10){\vector(0,+1){60}}  \put(+115,+70){$y$}
\put(+145,+40){\vector(0,+1){10}}
\put(+95,+40){\vector(0,-1){10}}
\put(+120,+65){\vector(-1,0){10}}
\put(+120,+15){\vector(+1,0){10}}

\put(+90,+70){$z = -0$}
\put(+105,+50){$e_{\phi}$}
\put(+130,+70){$A^{WY}_{\phi}= +g$}

\put(+90,+110){\vector(+1,0){60}}  \put(+150,+115){$x$}
\put(+120,+80){\vector(0,+1){60}}  \put(+115,+140){$y$}
\put(+145,+110){\vector(0,+1){10}}
\put(+95,+110){\vector(0,-1){10}}
\put(+120,+135){\vector(-1,0){10}}
\put(+120,+85){\vector(+1,0){10}}

\end{picture}

Certainly, such considerations are hinting and  intuitive  rather  than
exact and conclusively formulated arguments. Evidently,  that
the~whole situation would be more satisfactory if we could determine
a~mathematically more strict  characterization for measuring
certain  amount  of  singularity carrying by those monopole potentials.
It is understable that one should  expect  an~$U(1)$-invariant character
of that desireable measure of singularity.

So, the immediate task to solve is invention of a~certain mathematical
procedure  clarifying and rationalizing  this matter through some special
heuristic construction.   To begin with some summarizing steps --- one  may
list  all three gauges  through  the~following schematic graphs
(of which  exact  form does not matter to us, rather location of points of
discontinuity  is essential only)

\begin{center}
Fig.2
\end{center}
\unitlength=0.75 mm
\begin{picture}(160,70)(0,0)
\special{em:linewidth 0.4pt}
\linethickness{0.4pt}
\put(20,0){\vector(0,1){60}}
\put(20,30){\vector(1,0){60}}
\put(60,25){$\pi$}
\put(80,25){$\theta$}
\put(60,30){\circle*{1}}
\put(60,10){\circle*{1}}
\put(20,30){\circle*{1}}
\put(20,10){\circle*{1}}
\put(10,10){-2}
\put(25,50){$ A_{\phi}^{D(+)} \; =\; g \; (\cos\theta - 1) $}
\put(20,30){\line(2,-1){40}}
\put(100,0){\vector(0,1){60}}
\put(100,30){\vector(1,0){60}}
\put(100,50){\line(2,-1){40}}
\put(140,25){$\pi$}
\put(160,25){$\theta$}
\put(100,50){\circle*{1}}
\put(100,30){\circle*{1}}
\put(140,30){\circle*{1}}
\put(20,10){\circle*{1}}
\put(90,50){+2}
\put(110,50){$ A_{\phi}^{D(-)} \; =\; g \; (\cos\theta + 1) $}
\end{picture}

\begin{center}
Fig.3
\end{center}

\vspace{-15mm}
\unitlength=0.75 mm
\begin{picture}(160,80)(0,0)
\special{em:linewidth 0.4pt}
\linethickness{0.4pt}
\put(20,0){\vector(0,1){45}}
\put(20,30){\vector(1,0){60}}
\put(60,25){$\pi$}
\put(80,25){$\theta$}
\put(60,30){\circle*{1}}
\put(60,20){\circle*{1}}
\put(40,30){\circle*{1}}
\put(20,40){\circle*{1}}    \put(20,20){\circle*{1}}
\put(10,20){-1} \put(10,40){+1}
\put(25,50){$ A_{\phi}^{S} \; = \; g\; \cos\theta$}
\put(20,40){\line(2,-1){40}}
\put(100,0){\vector(0,1){45}}
\put(100,30){\vector(1,0){60}}
\put(100,30){\line(2,-1){20}}
\put(120,40){\line(2,-1){20}}
\put(140,25){$\pi$}
\put(160,25){$\theta$}
\put(100,30){\circle*{1}} \put(100,20){\circle*{1}} \put(100,40){\circle*{1}}
\put(120,20){\circle*{1}}
\put(120,40){\circle*{1}}
\put(140,30){\circle*{1}}
\put(90,20){-1} \put(90,40){+1}
\put(75,50){$  A_{\phi}^{WY} = \{ \; g \; (\cos\theta  - 1 ) \bigcup
                                    g \; ( \cos\theta + 1 )\;  \} $}
\end{picture}
\vspace{5mm}

Continuing this series of graphs, else one type of picture  may  be
naturally suggested (it might be called {\em anti}-(Wu-Yang) gauge)

\begin{center}
Fig.4
\end{center}

\unitlength=0.75 mm
\begin{picture}(160,60)(0,0)
\special{em:linewidth 0.4pt}
\linethickness{0.4pt}
\put(20,0){\vector(0,1){60}}
\put(20,30){\vector(1,0){50}}
\put(60,25){$\pi$}
\put(80,25){$\theta$}
\put(20,20){\circle*{1}} \put(20,40){\circle*{1}}
\put(40,20){\circle*{1}} \put(40,40){\circle*{1}}
\put(60,30){\circle*{1}}
\put(60,10){\circle*{1}}
\put(20,30){\circle*{1}}
\put(20,10){\circle*{1}}
\put(10,10){-2}  \put(10,20){-1}
\put(10,40){+1}  \put(10,50){+2}
\put(40,45){$A^{WY}_{anti} = \{ \; g\; (\cos \theta + 1) \cup
                                    g\; (\cos \theta - 1) \; \} $}
\put(20,50){\line(2,-1){20}}
\put(40,20){\line(2,-1){20}}
\end{picture}
\vspace{5mm}

Obviously, $U(1)$-gauge transformations  act  effectively  just
within a~fixed value of the parametre $g$, and all the  more  ---
the~separation  of  g's  into  positive  and  negative  ones  is  very
substantial (in the~above figures, the~positive values $g$'s are meant).

So, the~question of special interest   is  ---  how  would  one
substantiate  correctness  of  the~above  claim  that  all   three
potentials are equally singular ones? What could serve as a~measure for
proper  quantitative  evaluation of their singular  properties?

A~possible answer is almost evident at once. Indeed, after  all
the~above  steps and transformations, the~problem has been
effectivelly reduced to a~neat if not trivial  task:  namely,  for
exploration into the~singularities  one  should  compare  all
the~monopole functions $A_{\phi }$ (for definiteness we will discuss
the~case $g > 0$ ) with a~non-singular potential. It should be rememberred
that all those functions $f^{g}(\theta)$ ought to be contrasted with
the~situation free of singularity, i. e. when a~function $f^{0}(\theta)$
has  the~regular boundary  conditions:
$$
f^{0}(\theta = 0 ) \; = \; f^{0}(\theta = \pi ) \; =\; 0\; .
\eqno(4.1)
$$

\noindent The latter indicates that dealing with the~mathematical problem of
singularity, we should rather regard those two  values
$\theta  = 0$  and  $\theta =\pi $
of the~interval $[\; 0, \;\pi \; ]$    as identified ones\footnote{They
certainly represent different regions in the geometric 3-space
$x_{1} ,\; x_{2}, \;x_{3}$;  but  instead  we  mean
something very different: a~space of functions  with  specific  (null)
boundery properties and given at the~interval $[\; 0 , \; \pi \;]$,  which
admit  identification of its bounding points.}.

Thus, all  this may  be  reformulated   mathematically  as  follows:
{\em any  regular problem may be associated with a~space of continious functions
on that  interval. That is, every Abelian situation, not having  at all any
discontinuity at  the~$x_{3}$   axis,
can  be  associated  with  a~function $f^{0}(\theta )$  of  null boundary
conditions at this axis}
$$
\lim_{at \; x_{3}-axis} A^{Reg.}_{\phi } \; = \; 0 \;:
\qquad A^{Reg.}_{\phi } (\theta)  \; \rightarrow \;
f(0) = 0 \; , \;\; f(\pi) = 0  \; .
\eqno(4.2)
$$

\noindent Evidently, the~above statement reflects only the following requirement:
(of course, in its the~most simple and particular form):
{\em in any regular case, the~electromagnetic potential $A^{0}_{\phi }$  must
approach zero as we approach the $x_{3}$-axis (along any direction) }.

It is natural (if not obvious) the~further assertion:
{\em in any  irregular  case,  the  electromagnetic  potential $A_{\phi }$  may
not approach zero as we approach the $x_{3}$-axis}:
$$
\lim_{at \; x_{3}-axis} A^{Irreg.}_{\phi } \neq 0 \; : \qquad
A^{Irreg.}_{\phi } (\theta)  \; \rightarrow \;
f(0) \neq 0 \; , \;\; f(\pi) \neq 0  \; .
\eqno(4.3)
$$

One remark of principle must be given. Indeed,  if
the original electromagnetic potential
$A^{Reg.}_{\phi }$ has been previously submited to a~gauge transformation in
accordance with
$$
A^{Reg.}_{\phi } \; \rightarrow \; A'^{Reg.}_{\phi } \; = \;
A^{Reg.}_{\phi } \;-\; i {e\hbar \over e} \; S \; \partial _{\phi }\; S^{-1}
\eqno(4.4a)
$$

\noindent then  the~question  as to whether a~given  potential
$A_{\phi }$ is regular or irregular --- is to be  solved  in
a~different way: namely,
in any regular case, the~electromagnetic potential $A'^{0}_{\phi }$
must behave so that the~relation of the~form
$$
\lim_{at \; x_{3}-axis} \left [ \;  A'^{Reg.}_{\phi } \; + \;
i\; {e\hbar \over e} \; S \; \partial _{\phi } \; S^{-1} \; \right ] \; = \; 0
\eqno(4.4b)
$$

\noindent holds. In turn for all irregular cases this  relation
 must  be violated  (by difinition):
$$
\lim_{at \; x_{3}-axis} \left [\; A'^{Irreg.}_{\phi } \; + \;
i\; {e\hbar \over e} \; S \; \partial _{\phi } \;
S^{-1}\; \right ]\; \neq \; 0 \; .
\eqno(4.4c)
$$

Now, turning again to the~monopole case, one can  easily
realize that all the~monopole potentials, being listed  above,
have the~same one feature: each of them may be related to space
of functions given on the~interval
$[\; 0 , \;\pi \; ] $ (when the~point  $\;\theta = 0\;$ is identified with
$\theta = \pi\; $)  and
having only one point of discontinuity. Moreover, it is evident at
a~glance that the~intensity of the~relevant  discontinuity  remains
the~same as we go over from one potential to another, just varying  in
their  location. An additional remark should be given: significant as
the~locally achieved continuity according to Wu-Yang approach might seen,
it is not as important as  the~plain mathematical fact that to each of gauges
used there corresponds almost just the~same Fourier $[\;0,\; \pi\;]$ poblem
with only single point of  discontinuity.

This observation should be formalized in a~suitable notation.
To this end, one may take the~following definition for a~measure of
singularity (as will be seen, it is invariant under any gauge transformations):
$$
{\bf \mu}_{inv.} (A_{\phi }) = {\bf \mu}_{inv.} [f(\theta )] \; \stackrel{def}{=}\;
{1 \over 2} \; \left [\; f(x_{0}+0) \; - \; f(x_{0}-0) \; \right ] \;
\eqno(4.5)
$$

\noindent where $x_{0}$  denotes a~point of discontinuity,
and ${\bf \mu}_{inv.} (A_{\phi })$ designates
a~measure of singularity of the~potential $A_{\phi }$  at  such a~point.
Then, for the~monopole  potentials  described  above, we  will have
$$
\begin{array}{rlll}
S-gauge  :  &    f(\pi -0) = -g  \; ,  &   f(0+0) = +g \; , &
{\bf \mu}_{inv.} (A^{S}_{\phi })  = +\;g \;\;  ;  \\
D-gauge  :  &    f(\pi -0) = -2g \; ,  &   f(0+0) = 0  \; , &
       {\bf \mu}_{inv.} (A^{D}_{\phi })  = +\;g \;\; ; \\
D'-gauge :  &    f(\pi -0) = 0   \; ,  &   f(0+0) = +2g \;  ,  &
{\bf \mu}_{inv.} (A^{D'}_{\phi }) = +\;g \;\;  ;  \\
(WY)-gauge : &  f(\pi /2-0) = -g\; ,  &   f(\pi /2+0) = +g \;, &
{\bf \mu}_{inv.} (A^{WY}_{\phi }) = +\; g  \; .
\end{array}
$$

All the~above  types of discontinious functions in the~interval
$[\;0,\; \pi\;]$  come under the~eminent Dirichlet theorem's conditions:
 --- let us write out it in full.

{\bf The Dirichlet theorem}:

{\it If a~function $f(x)$ is given on segment $[-\pi, \pi]$, being bounded,
piecewise continuous and piecewise monotonic one, then its trigonometric
series converges at all the points of the segment. If $S(x)$ represents
a~sum of the trigonometric series for the function $F(x)$, then
at all the points
of continuity of this function, the equality $S(x)=F(x)$ holds;  whereas at
all points of discontinuity (there must exist just a~finit number of them)
one gets only}
$$
S(x) \; = \; {1 \over 2}\; \left [\; F(x-0) \;+\; F(x+0)\; \right]\; .
\eqno(4.6a)
$$

\noindent  {\it In addition, the identity}
$$
S(\pi) \; = \; S(-\pi) \; = \; {1 \over 2}\; [\; F(\pi -0) \;+\; F(\pi+0)\;]
\eqno(4.6b)
$$

 \noindent {\it holds}.

The above limitations on functions assumed in this theorem are often called
the~Dirichlet conditions. It should be amphasized that they
include essentially
both piecewise continuity and piecewise monotonity, and none of them cannot be
violated or waived. It is obvious that in (reasonable) physical applications,
likewize  in the~situation under consideration, these Dirichlet conditions
are likely to be satisfied.

Finally, turning to the {\em anti} Wu-Yang potential, we  notice
two points of  discontinuity: those  are
 $\theta  = 0(\pi )$  and $\theta = \pi /2$.
Taking in mind this example, a~more extendent definition for
$\mu ( A_{\phi })$ might be suggested:
$$
{\bf \mu}_{inv.} (A_{\phi }) = {\bf \mu}_{inv.}[f(\theta)]  \;
\stackrel{def}{=} \;
 \sum  {1 \over 2}\; \left [\; f(x_{i}+0) \; - \; f(x_{i}-0)\; \right ]
\eqno(4.7)
$$

\noindent where $x_{i}$ denotes all points of discontinuity (here there
 are  two ones). Thus, one has
$$
f(\pi /2-0)= + g \; , \;\; f(\pi /2+0) = - g \; , \;\;
 and \;\; f(\pi -0) = - 2g \; , \;\;  f(0+0) = + 2g
$$

\noindent and further
$$
{\bf \mu}_{inv.} (A^{anti(WY)}_{\phi }) \; =\;  {1 \over 2} \; g \;
\left  [\;  (-1-1)\;  +\; (2+2)\; \right ] = + \;g \; ;
$$

\noindent i. e. the~same value
${\bf \mu}_{inv.} (A^{anti(WY)}_{\phi }) = + g$ has
been found again.

The~latter example has been of unexpected interest because  it
shows some pecularity of the above  quantity $\mu (\vec{A})$. The matter is
that from intuitive viewpoint, the~case of the~potential
 $A^{anti(WY)}_{\phi }$
seems  much  more discontinuous in comparison with all three previous ones,
as it contains two singular points whereas each of the~others exhibit
only one point. However, the~result has turned out to be exactly
the~same. What is the~matter?

This  example  points to a~reasonable requirement for else one
additional characteristic to describe other sides of the~singularities
encountered above:
$$
{\bf \mu}_{addit.} (A_{\phi }) = {\bf \mu}_{addit.}[f(\theta)] \;
 \stackrel{def} {=} \;  \sum  {1 \over 2} \;
\mid f(x_{i}+0) \; - \; f(x_{i}-0) \mid    \; .
\eqno(4.8)
$$

\noindent Thus, one has
$$
{\bf \mu}_{addit.} (A_{\phi })^{anti-WY} = +4 \; g      \; .
\eqno(4.9)
$$

To clarify and spell out all the~sense of the two quantities
${\bf \mu}_{inv.}$  and  $ {\bf{\mu}}_{addit.}$,  let us introduce,
for heuristical purposes,
certain analogues  of  the~used above monopole gauges
for an~artificial situation when any
electromagnetic field (i. e. $\vec{E}$ and $\vec{B}$) vanish.
Those {\em imaginary}
electromagnetic fields may be represented just by certain {\em unphysical}
potentials. Those may be sketched  by  the~following  figures   (supposing
that for the~Schwinger's gauge, the~electromagnetic potential
vanishes:  $A_{\phi}^{(0)S} \equiv 0$)

\begin{center}
Fig.5
\end{center}

\vspace{5mm}
\unitlength=0.75 mm
\begin{picture}(160,60)(0,0)
\special{em:linewidth 0.4pt}
\linethickness{0.4pt}
\put(20,0){\vector(0,1){60}}
\put(20,30){\vector(1,0){60}}
\put(60,25){$\pi$}
\put(80,25){$\theta$}
\put(20,20){\circle*{1}} \put(20,40){\circle*{1}}
\put(40,20){\circle*{1}} \put(40,40){\circle*{1}}
\put(60,40){\circle*{1}} \put(20,30){\circle*{1}}
\put(60,30){\circle*{1}}
\put(10,20){-1}  \put(10,40){+1}
\put(40,50){$A_{\phi}^{(0)WY} = \{ - g  \cup  + g \}$}
\put(20,20){\line(1,0){20}} \put(40,40){\line(1,0){20}}
\put(120,0){\vector(0,1){60}}  \put(120,30){\vector(1,0){60}}
\put(160,25){$\pi$}            \put(180,25){$\theta$}
\put(120,20){\circle*{1}}      \put(120,40){\circle*{1}}
\put(160,40){\circle*{1}}      \put(160,20){\circle*{1}}
\put(160,40){\circle*{1}}      \put(120,30){\circle*{1}}
\put(160,30){\circle*{1}}
\put(110,20){-1}               \put(110,40){+1}
\put(140,50){$A_{\phi}^{(0)D} =  +g  \; or \; -g $}
\put(120,20){\line(1,0){40}}   \put(120,40){\line(1,0){40}}
\end{picture}
\vspace{5mm}

\noindent Correspondingly, one has
$$
\mu_{inv.}(A_{\phi}^{(0)WY}) = g [\; ( \; f(0+0) - f(0-0)\;) \;+\;
(\; f(\pi/2 + 0) - f(\pi/2 + 0)\;) \;] =
$$
$$
g \; [-1-1+1+1] = 0 \;\; , \;\; and \;\;
\mu_{addit.}(A_{\phi}^{(0)WY}) = +4 \; g \;
$$

\noindent those two relations can be interpreted
 as follows:
$\;(A)\;\;$
$\mu_{inv.}(A_{\phi}^{(0)WY}) = 0$ points to the~absence of any
{\em real} singularity at vacuum-like state of electromagnetic field, though
$A_{\phi}^{(0)WY}$  is not null; $\;(B)\;\;
\mu_{addit.}(A_{\phi}^{(0)WY}) \neq 0$ proves
$\mu_{addit.}$ as a~characteristic of singularity properties of gauge
transformations involved here.

In case of $A_{\phi}^{(0)D}$, two relations
$$
\mu_{inv.}(A_{\phi}^{(0)D}) = (g \;- \;g) = 0 \; , \qquad
\mu_{addit.}(A_{\phi}^{(0)D}) = \mid g - g \mid = 0 \; .
$$

\noindent may be interpreted in a~similar way:
$\mu_{inv.}(A_{\phi}^{(0)D}) = 0$ conforms to $\;(A)\;\;$  above;
and $\mu_{addit.}(A_{\phi}^{(0)D}) = 0$ corresponds with that
$A_{\phi}^{(0)D}$ has no singularity in continuity properties at
the interval $[0, \; \pi]$ (thereby, it  is in accoradance with
$\;(B)\;\;$ above.

\subsection*{5. On singulatities, Aharonov-Bohm effects and some inherent
     requirements implied by the~conventional gauge principle}

Some immediate consequences of the~above constracted two mathematical
quantities $\mu_{inv.}$ and $\mu_{addit.}$ might be of
noticeable utility in
quite other physical phenomena of much more generalized nature.  At
that point, we are going to pass away from the~monopole matter and to
deal with certain aspects  of the~well known, and extensively learnt
in the~literature, Aharonov-Bohm effect and will discuss it
in somewhat new terms.  Similtaneously we shall touch on
the~conventional gauge principle's inherent structure.

Results obtained in that way, though in certain their sides are not
without lacking in rigour, seems attractive and
quite plausible. In any case, those  developments hold promise of appreciable
progress in clearing up, even if not explaining away completely, and
definitive resolving these long standing paradoxical phenomena on
notoriously known and predicted as physically observable manifestations
of unphysical and subsidiary field $A^{0}_{\alpha}(x)$ related to vanishing
field  $(\vec{E}, \vec{B})$ ). Just such a particular aspect of
the~whole much more comprehensive  matter of Aharonov-Bohm effect
will be meant in the~following. In turn, on that line of
arguments, a~specific view on various monopole gauges will be worked
out\footnote{To avoid misunderstanding it should
be stressed that the subsequent part of the work bears character of
discussion rather than strict conclusive results.}.

Let us begin from the~very generalities. So, it must be accepted that
under all circumstances any entity, if it is considered just as
a~mathematical construct but not existing in reality, should not be
a~sourse of tension and contradiction even at the~level of
theoretical arguments or mental experiments. If the inverse arises
(as it is so now) then, in the~first instance, one ought to take
notice of a~possibly wrong unadequate understanding or interpreting
of the~situation and hence to look into, in the~first, place just
those aspecs of the~problem, rather than to bring out, somewhat
routinely, any new confirmations to such an~already fixed paradox.

To clarify more exactly what I mean here, let us look at just one
side of the~matter. That is the~following: an~arbitrary $U(1)$-gauge
transformation --- at the~level of both the~electromagnetic
$4$-potential  $A_{\alpha}(x)$ and the~wave functions for a~quantum
mechanical particles placed in the~field of a~magnetic charge ---
as a matter of fact
carries always a~certain amount of irregularity (or may be better to
say --- {\em singularity}; the~latter term itself might be easily
extended so that to  cover all the~changes produced by those gauge
transformations). In other words, any explicitly given picture of
a~chosen  physical  system always bears a~mark which is in exact
correspondance with a~respective gauge.

Unfortunately, among physisists, mainly an~idea of gauge invariance
itself has been fixed in mind  --- so strongly that, usually, they
pass over some its inherent peculiarities and  subtleties which
accompany this undoubtly grand (mathematicaly and physicaly)
structure. In particular, the~substantial affecting  of
the~relevant representing picture of a~physical system (i.e.
an~alteration of regularity properties), though being accepted and
recognized as such, seems often less significant.  As a~sequel, in
majority of cases, they incline to detract from the~necessity of
the~accurate following it,  so that often this alteration turns out
to be not remembered at all.  But from this attitude only one step
remains to face (unexpectedly just at first  sight) the~paradox on
physical manifestations of not existing fields and further ---
a~variety  of Aharonov-Bohm effects.

Evidently,
the~above question  of whether the~different gauges (Dirac, Schwinger,
Wu-Yang's) are just different representatives for a~unique physical object
or not --- may be qualified as beloning to the~same problems which
exist and manifest themselves  yet in situations with no magnetic charge.
In this connection, the~very general view might be claimed that nature hardly
produces so extensively really different physical objects,
which originate from one that and which can be obtained trough
such  {\em a~non-trivial} exploiting of the~$U(1)$ gauge transformations.
Instead I think one should have worked out such  a~viewpoint that
could guarantee any gauge transformation will not be a~thing producing
physical effects\footnote{If the~inverse  were true, fantasy and
ingenuity of nature would  exceed any reasonable limits
beyond our expectations, to say the~least.}.

So, our immediate concern is the~question --- how one should reflect and act
in order to select {\em a~proper representative}
 $A_{\alpha}^{proper}$ from the~whole set of possible candidates
$$
 \left \{ A_{\alpha}^{proper} \; - \; i {\hbar c \over e}\; U(x) \;
                    {\partial \over \partial x^{\alpha}}\; U(x) \; ; \;
\qquad U(x) \in U(1)_{loc.} \right \}  \;  .
\eqno(5.1)
$$

But it is the~moment to remind that differences between all
elements of this set (5.1) are in~evidence and they can be  imediately seen:
those  are their boundary condition properties (or in other terms,
their singularities).
The~reasons for passing over, generaly, such peculiarities can be quite
easily understood. Seemingly those are:  first, the~autority of
 a~gauge  invariance principle itself; second, familiar and imbibed
from the~very begining, the~quantum-mechanical interpretation of a~square
modulus $\mid \Psi(x) \mid ^{2}$.  These both lie equally
at the~bottom of our understating and even ignoring such  {\em minor
alterations} in boundery properties.
Especially  the second one detracts  from the~importance of those subtleties.
In contrust, further I shall suppose that the~true lies  within just
those boundary condition alterations.

Just  in this point, the above introduced quantities
$\mu_{inv.}(A_{\alpha})$ and $\mu_{addit.}(A_{\alpha})$ have their
practical side\footnote{It is  rather surprising that the~technique, based on
the~use of them, turns out to be geared in such a~perfect fashion to
handling this matter.}.
The first measure $\mu_{inv.}(A_{\alpha}) $ provides us with a~general
characteristic for the~whole  class
$$
\left \{\; A_{\alpha}^{proper} \; - \; i {\hbar c \over e} U(x) \;
        {\partial \over \partial x^{\alpha}} U(x) \; ; \;
\Psi'(x) = U(x) \Psi(x) ; \;
 U(x) \in U(1)_{loc.} \; \right \}
\eqno(5.2)
$$

\noindent  the $\mu_{inv.}(A_{\alpha}) $  remains the~same for all
those electromagnetic potentials related to each other by means of any
gauge transformations (one should remember  those may be  piecewize
continuous as well as monotonic ones).
In other words, this measure $\mu_{inv.}(A_{\alpha})$ cannot be changed
by the~use of any, even with some special purposes constructed,
singular gauge transformations; therefore, this quantity can serve
the~inherent characteristic of physical system itself.

In contrast to this,  the~second measure $\mu_{addit.}(A_{\alpha})$
generally changes when passing from one gauge to
another just through the~use of any irregular gauge transformations.
Therefore such a~measure can serve to trace the~use of any singular
(piecewise continuous) gauge transformations.

By the way, a quite determined hierarchy between them (measures)
might be presuposed in advance: in every separate example of a~physical
system,  after we have calculated its concomitant quantity
$\mu_{inv.}(A_{\alpha}) $ and then
set all  various  gauges  into correspondence with the values of
the measure   $\mu_{addit.}(A_{\alpha})$, we may expect that the~set
$$
\left \{\; \mu_{addit.}(A_{\alpha}^{U} )\; ; \;\; U \in U(1) \; \right \}
$$

\noindent implies a~certain  minimum value; the lowest bounding evidently
exists --- this seems more than plausible, and  further it should coincide
with the~value of first (invariant) measure.
However, this lowest bounding  value of $\mu_{addit.}(A_{\alpha})$
yet does not  set just one basis
apart from all the~others, it  provides us only with a~ set of candidates to
preferable one.  All these candidates can be reffered to each other by means
of gauge transformations;  and what is more, such transformations
are formed by either piecewize continuous
or globally continuous functions of 3-space coordinates only;
and presupposedly never they are continuous functions of  3-space
geometric points.

{\em Under all these circumstances, the~single  and only way out may be
put forward: namely, that
a~preferable basis will be discovered if there  exist indeed
a~continuous function of 3-space points (the latter merely
could be affected and even destroyed by the~use of a~singular gauge
transformation).
Thus,  the deciding (and essentially only remaining) step in searching
a~preferable
gauge consist in the~following: one ouhgt to find a~gauge without
singularity}\footnote{Remember that here, for the~moment,
only a~situation free of any magnetic
charge or any other singular cases, is  discussed.}.

In other words, looking at the~behaviour of relevant potentials or wave
functions we would find that those are single-valued functions
 of 3-space geometric points just in a~unique basis. This conjecture
(and conclusion)  seems quite justified.
Moreover, this assumption on existence  of a~gauge with its concomitant
single-valued  electromagnetic potentials and likewize single valued
 wave functions, seems inevitable and even very
desirable; otherwise the concept of single-valued physical fields (potentials,
wave  functions, and so on) even teoretically cannot ve discussed.
The problem in issue can be reformulated differently:
either we manage to arrive at a deternination of a~gauge being {\em better}
than all other or inevitably  we have to reconcile ourselves to a~variety of
physical predictions where just one  that would be desirable. No other way out
exists.

In this connection, one should take special notice of the fact that such
an~indeterminacy substantially touches the physical gauge principle itself.
Indeed, the~situation which we face here may be sketched as follows
$$
if \;\; g = 0 \; :
$$
$$
\{`short' \; gauge \; principle\; \oplus \;\;
a \;preferable \; basis \} \; = \; GAUGE \; PRINCIPLE \;
$$
$$
if \;\; g \neq 0  \; :
$$
$$
\{`short' \; gauge \; principle\;  \oplus \;
no \; \; preferable \;  basis \} \; = \;
What \;\;is \;\;  this ?
$$

But not having any preferable basis, in the~second case, what is
the~meaning of the~{\em physical identifying} of all the~pictures as
describing the~same one situation though in various  gauges?
The~mathematical situation
under consideration (specified by  allowance of any not removable
singularities) does not provide us with a~sufficiently justified criterion
about a~certain physically invariant essence which just can be described
in many ways through the~use of various gauges.
 So,  in such a~new and at the~first sight only  slightly altered situation,
any counterpart of the~second essential
constituent (see above) in the~ordinary gauge-invariance principle cannot be
brought out. For these reasons, in my opinion, the~case in issue
should be associated with a~double (two-faced) status for  symmetry
 transformations of the~group (here)
$U(1)_{loc.}$  rather  then conventional unique  one:
$$
G = U(1)_{anti-gauge} \otimes U(1)_{gauge} \;\;\;  instead \;\;\;
 G_{conventinal} =  U(1)_{gauge}  \;  .
\eqno(5.3)
$$

\noindent
It is no use blinking at the fact that after any (not removed) singularities
had been allowed in physics, then all our subsequent  attempts to retain
the~conventional gauge invariance principle unchanged as such were doomed
to failure.  In other words, one may say that  the `innocent' allowance
itself of sustantially  singular potentials (which cannot be approached by
single-valued functions of $3$-space geometric points) turns out to be utterly
destructive to the~essence of old and standart gauge-invariance principle, i.e.
translating it into something  totaly different.
Thus, either one should reject
the~singularities (the~simplest representative of wich is
$\mu_{inv.}(\vec{A}_{\alpha}(x)) \neq 0$)
considering  them as unphysical ones, or one ouhgt to accept physics with
an $anti-gauge\;\;\otimes\;\; gauge\;\; symmetry$  as above in (5.3).

In other words, the view point may be advanced
(this claim touches certain  sides of  the~ Aharonov-Bohm
phenomena too) that gauge choice-based
paradoxes should be regarded in some extent  as
a misunderstanding  arisen out of not sufficiently exact and
elaborated terminology  rather than from any actual contradistinction
within  the~physical theory itself. As a matter of fact, just making
the~terminology used more accurate and precise might
axplain away paradoxical aspects of all effects of that kind.

\subsection*{6. Metodology conclusions and discussion}

Thus,  the  article  has  brought  together  such  apparently
unrelated ideas as to the monopole charge and old conventional
Fourier series analysis; in particular, the Dirichlet theorem seems
most significant in this connection. In the~author opinion, the plain
and striking fact is that the~monopole situation  entirely comes
under this purely mathematical theory with many its concomitant
subtleties of both mathematics and physics.  So, just  going back to
some classical fundamentals of this theory leads us, as might be
hoped, to a~constructive reformulation of certain purely physical
monopole problems.  Such  a~synthesis  is  always attractive to
theoreticians; the more so as the new insight gained holds promise
of  further progress,  which  could  be  of  great importance in, for
example, understanding  the  interrelationship among singularities
(in particular,  monopole's), the concept of single-valuedness of
wave functions, and gauge principle. Even if this last hope is not
fulfilled, or outcomes  achieved turn out to be less than expected,
one will  learn more about the nature of monopoles and how they
should be rationalized so  as to judge its further  evaluation.  In
any case some interesting lessons  can be lernt from the above
suggrsted approach to the monopole mattter,  which might  save
us from having unjustified expectations and from dwelling too much
on eliminations of the effects of singularities. that connot
be removed in principle.

In addition, ome may note that
the  intrinsic potential  power  of  the  approach  based on
the~Fourie  analysis is that no assumption regarding the
nature of the any underlying equations are necessary, so that
various physical  systems  are automatically included.  Definitely,
the above trearting, while setting a~pattern for possible
considerations on that line, admits further extening and developing
to  various situations where some singular potentials could occur. In
so doing, certainly, the arrived generalizations can considerably
differ from the present variant in appearance, definitions,
terminology, and else in a number of characteristics, but the basic
spirit seemingly is going to be the same: namely, that the
single-valuedness of potentials and wave functions with respect to
3-space geometric points and requirement to trace accurately where
and how this property is modified, both are to be considered as a
basis for any assessing extent of singularity of every situation as
well as variuos gauges transformations.

Some more practical sides of the~monopole's
presence concerning peculiarities of wave
functions of quantum-mechanical particles affected by the monopole
potential will be considered in a sepatate paper. Here I only wish
 note that  the features  of the
$S$-, $D$-, ($W-Y$)-gauges find their natural corollary  in  boundary
condition properties of corresponding wave functions. In
particularly, the common argument to   justify applying    the
$(W-Y$)-approach to quantum-mechanical particle-monopole problem is
that it allows us to get rid  of discontinuity of wave function (at
$x_{3}$  axis).  In the same time, they pass over the  fact  that
some  discontinuity appears at $x_{3} = 0$  plane. More exactly,
they  say  that  in  the region of overlapping, the functions $\Psi
^{(N)}(x)$   and $\Psi ^{(S)}(x)$ vary in phase factor $\exp
 (2ieg/h)$, which is not essential   to  any physically observable
quantities. However, it should be stressed that particle wave
functions in the  $S$- and  $D$-gauges have a~very special violation,
namely, both of them  undergo  exponential kinds of discontinuity,
which are, by the same token,  irrelevant to physically observable
quantities as if we have utilized the $W-Y$ gauge .


\end{document}